\documentclass[aps,prl,twocolumn]{revtex4}
\usepackage{graphicx}
\newcommand\beq{\begin{equation}}
\newcommand\eeq{\end{equation}}
\newcommand\bear{\begin{eqnarray}}
\newcommand\eear{\end{eqnarray}}

\begin{document}

\title{Effect of Protonation on the electronic properties of DNA base pairs: 
Applications for molecular electronics}

\author{Sairam S. Mallajosyula and Swapan K Pati}
\affiliation{Theoretical Sciences Unit,
Jawaharlal Nehru Center for Advanced Scientific Research,
Jakkur Campus, Bangalore 560 064, India.}

\date{\today}

\begin{abstract}

{Protonation of DNA basepairs is a reversible phenomenon which can be 
controlled by tuning the pH of the system. Under mild acidic conditions, 
the hydrogen bonding pattern of the DNA basepairs undergoes a change. We study 
the effect of protonation on the electronic properties of the DNA 
basepairs to probe for possible molecular electronics applications. We 
find that, under mild acidic pH conditions, the A:T basepair shows excellent 
rectification behaviour which is, however, absent in the G:C basepair. The mechanism 
of rectification has been discussed using a simple chemical potential model. 
We also consider the non-canonical A:A basepair and find that it can be used as 
efficient pH dependent molecular switch. The switching action 
in A:A basepair is explained in the light of $\pi$-$\pi$ 
interactions which lead to efficient delocalization over the entire basepair.}

\end{abstract}
\maketitle

The research on electron transport through molecules is gaining a lot of 
attention in recent years due to their potential applications in a large 
class of nanoscale technologies\cite{review}. With advances in 
nanotechnology, there has been an active effort to use biomolecules for 
molecular electronic applications. This has been especially true for DNA 
based electronics owing to the unique geometry and stability of DNA which 
is built up on weak intermolecular $\pi$-stacking and specific hydrogen 
bonding (H-bonding) interactions between Adenine:Thymine (A:T) and 
Guanine:Cytosine (G:C) basepairs\cite{DNA_initial}.  

So far, the studies on DNA conductivity have been mostly restricted to understanding 
the effect of $\pi$-$\pi$ interactions between the DNA bases. This is due to 
the fact that experimental work on DNA conductance is carried out with DNA 
strands and the measurements are always between end to end of the DNA strand, 
or, in other words, about the helical axis of the DNA strand\cite{DNA_reviews}. 
The importance of H-Bonding in these experiments have been restricted to the 
formation of the DNA double helix, which stems from the specific basepairing 
pattern among the DNA bases.

In this work, we have analyzed the role played by the H-bonding between the DNA 
bases under the influence of mild acidic conditions. It is known that the 
protonation of the DNA basepairs under mild acidic conditions is a 
reversible phenomenon and that it disrupts the H-Bonding profile between 
the DNA bases\cite{DNA_structure}. We find that, this protonation under 
mild conditions changes the electronic properties of the DNA basepairs 
significantly and the effect being different for different basepairs. The 
effects of pH dependent reversible protonation-deprotonation on the 
electronic properties of a molecular system have been recently studied 
experimentally for di-block molecular diodes, where the inversion of the 
rectifying effect was found on selective protonation and deprotonation
\cite{JACS_diode}. 

We have used the SIESTA and Gaussian 03 codes for our first principles 
DFT simulations\cite{siesta,g03}. Within SIESTA, we have used the 
generalized gradient approximation for the exchange-correlation energy 
functional in the version of Perdew, Burke, and Ernzerhof\cite{GGA}. 
A double-$\zeta$ basis set with the polarization orbitals has been 
included for all the atoms\cite{sankey}. To model the gold electrodes, we 
placed two 27 atoms gold clusters at either end of the DNA basepairs. 
The coordinates of the gold clusters were obtained by relaxing two 
Au(111) surfaces separated by 30 $\AA$, in a periodically repeated 
cell. Each surface was comprised of three Au(111) planes in a 2x2 
surface unit cell along the [111] direction. The clusters thus derived 
from the relaxed surface calculations were then attached to the DNA 
basepairs via a thiol linker (-CH$_2$S). The Sulphur head group of 
the thiol linker was aligned on the fcc hollow site of the Au(111) 
surface using S-Au distance of 2.51 $\AA$, previously reported in 
literature \cite{Rosa_surface}. For the subsequent SIESTA calculations, 
the initial geometries of the basepairs were generated by Gaussian 03 
calculations. Within Gaussian 03, we have used the polarizable continuum 
model with water dielectric to model the solvation effects. The large 
dielectric continuum of water reduces the electrostatic repulsions 
and stabilizes the formal positive charges introduced due to protonation. 
We have used the LANL2DZ basis set for obtaining the structures of the 
basepairs constrained between the two gold electrodes, wherein the gold 
electrodes are modeled as an equilateral triangle with an Au-Au distance 
of 2.9 $\AA$ which was obtained from the relaxed gold surfaces and which 
matches well with the previously reported experimental results\cite{lanl}. All the 
calculations in SIESTA for the DNA basepairs attached to the gold 
clusters have been performed in a 30 x 30 x 60 $\AA^3$ supercell, which 
is large enough for interactions between the neighboring fragments to 
be negligible. To model the reversible denaturation of the DNA basepairs, 
which can be controlled by pH variation, we have considered only the 
protonation of the amino nitrogen's, which can be easily achieved under 
mild acidic conditions\cite{DNA_prot}.

Using the above described methodology, we have studied three different basepairs, 
namely, A:T (Adenine:Thymine), G:C (Guanine:Cytosine), and A:A (Adenine:
Adenine). The first two are the naturally occurring watson-crick basepairs 
where the number of H-bonds between the bases are two for A:T and three 
for G:C. As a first observation we note that on protonation both Adenine
:Thymine (A:T) and Guanine:Cytosine (G:C) basepairs lose the N-H...N 
H-bond, however, they still retain the N-H...O H-bond. Thus, for a 
complete loss of H-bonding under mild acidic conditions, we consider 
the non-canonical Adenine:Adenine (A:A) basepair, which loses both the 
N-H...N H-bonds on protonation. Here, we would like to point out that 
the systems studied by us can be realized by experimental synthesis. 
We give a probable scheme for the experimental realization in Fig.1(a) based 
on the pioneering work of Mirkin {\it et al.} and Alivisatos {\it et al.} 
\cite{mirkin_and_Alivisatos}. The first step in the experimental method involves 
attaching DNA bases capped with thiol groups, which bind to gold, to a gold surface. 
Then the complementary bases are similarly attached to the surface of a gold 
nanocluster {\it via} thiol linkages. The bases attached to the gold nanoclustres are 
then immobilized on the gold surface by spotting solutions. The immobilization is 
governed by the complementary base pairing. For the synthesis of the non-canonical 
basepair A:A, one needs to functionalize both the gold surface and gold cluster with 
Adenine. Hereafter, we denote the non-protonated basepairs as A:T$_{NP}$, G:C
$_{NP}$, A:A$_{NP}$ and the protonated basepairs as A:T$_{P}$, G:C
$_{P}$, A:A$_{P}$. 

\begin{figure}
\includegraphics[scale=0.08,angle=0]{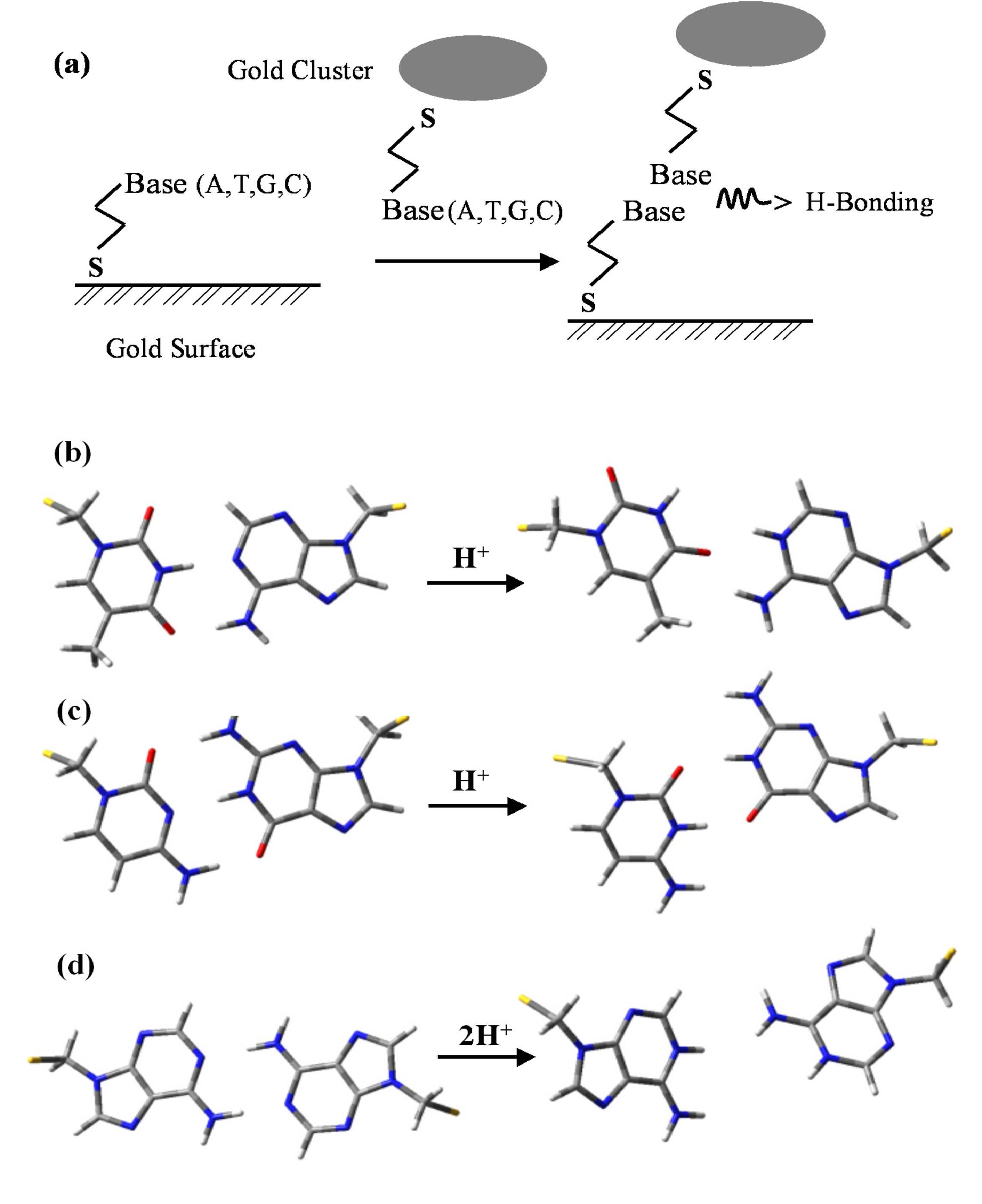}
\caption{(color online) (a) Probable scheme for experimental realization. Optimized 
geometries of the DNA basepairs: (b) A:T$_{NP}$, A:T$_{P}$, (c) G:C
$_{NP}$, G:C$_{P}$, and (d) A:A$_{NP}$, A:A$_{P}$. The Gold clusters 
have been omitted for clarity.}
\end{figure}

Upon optimization of the nonprotonated and the protonated basepairs, 
we find that the basepairs do not get aligned perpendicular to the 
surface of the gold cluster. The tilt with respect to the surface is 
about 106$^{\circ}$ and this is due to the presence of the thiol linker 
which are used to attach the basepairs to the gold cluster surface. 
In Fig.1(b), (c) and (d), we present the optimized geometries of 
A:T$_{NP,P}$, G:C$_{NP,P}$ and A:A$_{NP,P}$, respectively. We find 
that the A:T$_{NP}$, G:C$_{NP}$ and A:A$_{NP}$ form planar structures 
due to almost linear H-bonding. However,  the planarity of the H-bonds 
are lost on protonation in A:T$_{P}$, G:C$_{P}$ and A:A$_{P}$. 

For A:T$_{P}$ (Fig.1(b)), we find that the protonation leads to the 
formation of a bidentate H-bond between two N-H and one O (1.82 $\AA$, 
$145^{\circ}$ and 1.92 $\AA$, $141^{\circ}$) as compared to the 
linear N-H...O (1.72 \AA, $173^{\circ}$) and N...H-N (1.67 \AA, 
$178^{\circ}$) H-bonds in A:T$_{NP}$. For G:C$_{P}$ (Fig.1(c)), the 
protonation disrupts the central N-H...N H-bond and leads 
to the reorientation of the bases to form two N-H...O (2.02 $\AA$, 
$174^{\circ}$ and 1.67$\AA$, $157^{\circ}$) H-bonds as compared to the 
two N-H...O (1.86 $\AA$, $179^{\circ}$ and 1.62$\AA$, $179^{\circ}$) 
H-bonds and one N-H..N (1.84 $\AA$, $178^{\circ}$) H-bond in G:C$_{NP}$. 
The case of A:A$_{P}$ (Fig.1(d)) is distinct. The protonation in this 
case disrupts the two N-H...N H-bonds and the resultant structure is 
stabilized by partial $\pi$-$\pi$ interactions between the Adenine bases. 
In A:A$_{NP}$, there are two equivalent N-H...N (1.84 $\AA$, $178^{\circ}$) 
H-bonds. Thus, the main point of our findings is that the protonation 
affects the nature and the 
strength of the H-bonds which leads to a considerable change in the 
structure of the basepairs with little change in energy as the N-H...N 
H-bond has the lowest H-Bond energy. We investigate the transport 
phenomenon of the non-protonated and the protonated basepairs to 
analyze the role played by H-bonds and the $\pi$-$\pi$ interactions affecting 
the electronic structure of the basepairs thereby the related conductivity
characteristics. 

To understand the transport properties of the systems under study, 
we have calculated the transmission function T(E) at different bias 
voltages using the Green's function methodology\cite{Datta}. The 
Hamiltonian (H) and the overlap (S) matrices are obtained from the 
self-consistent DFT calculations. To model the effect of bulk 
electrodes, we modify the Hamiltonian of the system by adding the 
imaginary self energy terms, $\Sigma_{L,R}$, (L and R being left 
and right electrode indices respectively) arising due to the bulk 
gold, to the orbitals of the gold cluster. Thus, the original 
Hamiltonian H gets modified to $\overline{H}$ as
\bear
\overline{H} = H+\Sigma_{L}+\Sigma_{R}
\eear
With the modified Hamiltonian, we calculate the Green's function 
G(E) and the transmission T(E) is then calculated as 
\bear
T(E)= Tr[\Gamma_{L}G(E)\Gamma_{R}G^{+}(E)] 
\eear
where $\Gamma_{L,R}$ are the anti hermetian parts of the self-energy 
matrices which describe the broadening of the energy levels of the 
system due to the coupling to the electrodes.
\bear
\Gamma_{L,R}(E) = i(\Sigma_{L,R}-\Sigma_{L,R}^{+})
\eear

\begin{figure}
\includegraphics[scale=0.08,angle=0]{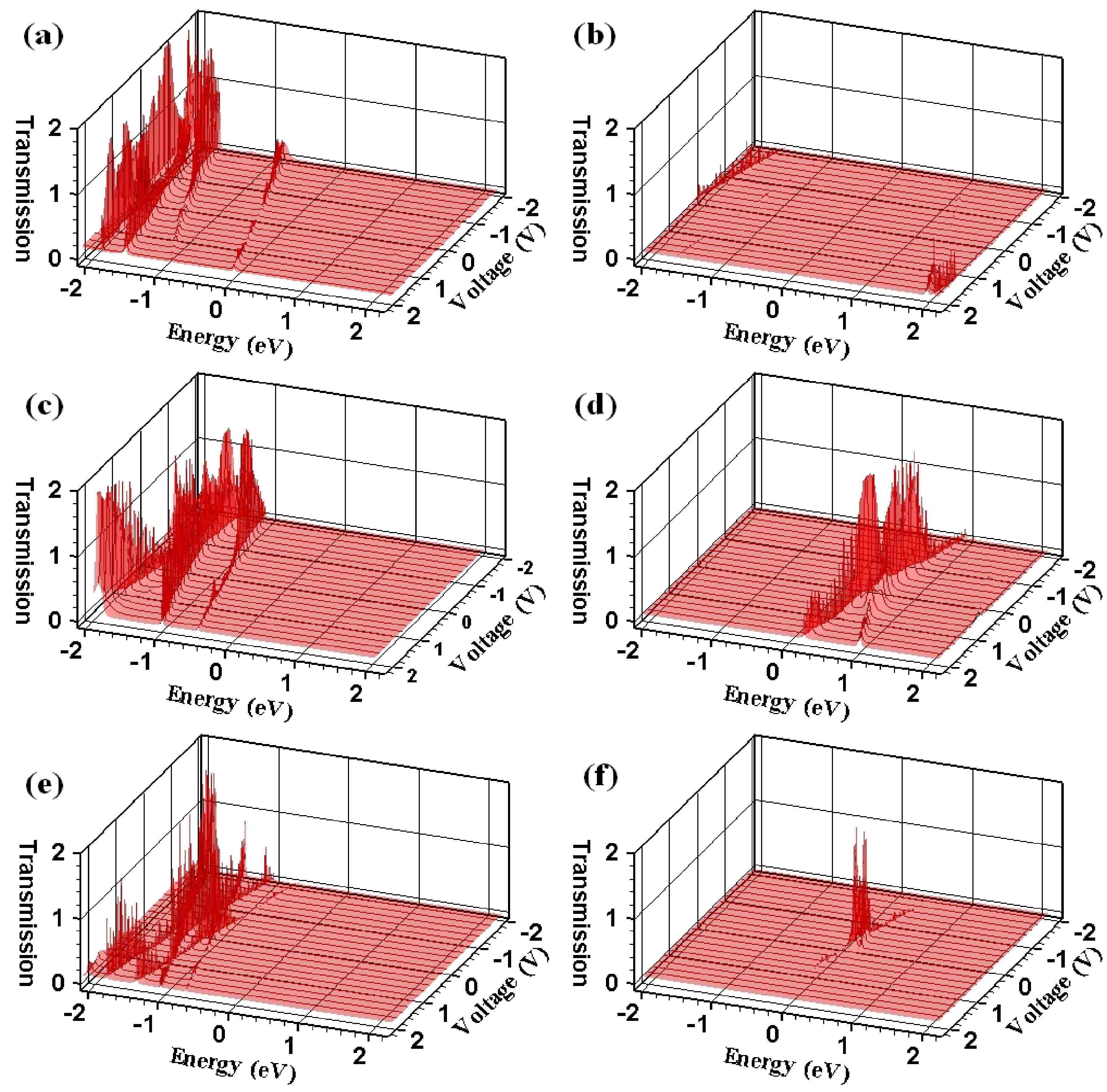}
\caption{(Color online) Transmission T(E) as function of applied bias voltage. 
(a) G:C$_{NP}$, (b) G:C$_{P}$, (c) A:T$_{NP}$, (d) A:T$_{P}$, 
(e) A:A$_{NP}$ and (f) A:A$_{P}$.}
\end{figure}

In Fig.2, we plot the transmission T(E) as a function of bias voltage 
for G:C$_{(NP,P)}$, A:T$_{(NP,P)}$ and A:A$_{(NP,P)}$. We 
restrict our discussion to the low-energy molecular orbitals which govern 
the transport phenomenon around the Fermi level ($E_{F}$). Note that, 
since we have incorporated a large number of Au atoms to model the extended 
molecule, the relative location of the molecular levels with respect 
to $E_{F}$ is accurately estimated. On examining the T(E) for all the 
cases in the energy range $E_{F}$-2 eV to $E_{F}$+2 eV, we find that 
the transmission for the non-protonated basepairs is governed by the 
occupied molecular orbitals whereas for the protonated basepairs, the 
transmission is controlled by the unoccupied molecular orbitals. We 
note that this is a direct consequence of the positive AEA (Adiabatic 
Electron Affinity) associated with the free bases\cite{sairam_syn}. 

\begin{figure}
\includegraphics[scale=0.08,angle=0]{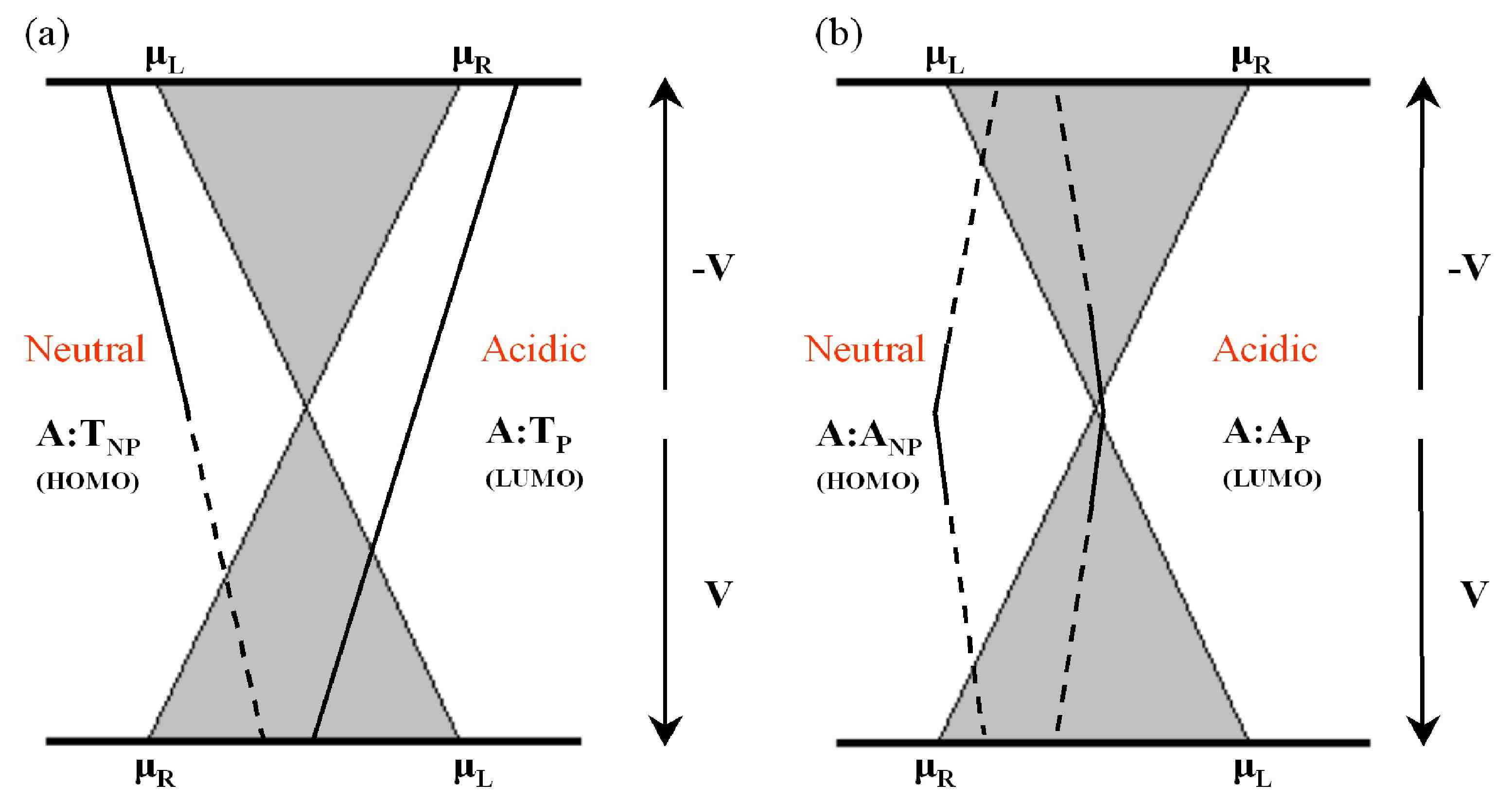}
\caption{(color online) (a) Rectification mechanism for A:T$_{P}$. (b) Switching action in 
noncanonical basepair A:A. Shaded portion illustrates the integrating region. 
Arrows indicate the bias direction. Solid lines indicate strong transmission and 
dotted line indicate weak transmission.}
\end{figure}

Before analyzing the T(E) data for the individual cases we point out the 
general features in the T(E) data plotted in Fig.2 . The naturally occurring Watson
-Crick basepairs A:T and G:C have an inherent asymmetry due to which we 
find that the transmission peaks follow the changes in the chemical 
potential of one electrode. Whereas, for the symmetric basepair A:A, we 
find no such preference and the transmission peaks follow a symmetric 
trend in both the forward and reverse bias directions. Similar shifts of 
molecular levels under the influence of an external bias have been studied 
before for a symmetrically and asymmetrically contacted molecule\cite{Taylor}.

The HOMO band of G:C$_{NP}$ (Fig.2(a)) which is 0.35 eV below the 
$E_{F}$ shows very weak zero bias transmission and it does not improve 
with a change in the bias voltage. However, the HOMO-1 band which is 1.35 eV 
below the $E_{F}$, shows strong transmission in the negative bias region, 
but not in the forward bias region. Also we note that in the negative 
bias region, the transmission peak moves away from the $E_{F}$ and thus 
we do not expect this transmission band to enter the integration region 
and have any pronounced effect on the Current-Voltage (I-V) characteristics 
of the system. For G:C$_{P}$ (Fig.2(b)), we do not find any strong transmission 
peaks in the energy range considered by us. Thus, our calculations show that the 
G:C$_{NP}$ and G:C$_{P}$ do not posses any features of interest for 
molecular electronics applications.  

However, the situation is very different for the A:T systems. As can be seen
from Fig.2c, the HOMO band for A:T$_{NP}$, which is 0.75 eV below the $E_{F}$, 
shows strong transmission in the negative bias region, but not in the 
forward bias region. However, due to the asymmetric nature of the electron 
distribution in the basepair, the transmission peak moves away from the 
$E_{F}$ with increase in bias, in the negative bias region. Thus, we do not 
expect this 
transmission band to enter the integration region to generate any nonzero 
I-V characteristics for small reverse bias. However, for the A:T$_{P}$ (Fig.2(d)), 
we observe 
that the LUMO band which is 0.47 eV above the 
$E_{F}$ shows very strong transmission at zero bias and appreciable 
transmission in the negative bias as well as in the forward bias region. We also 
find that in the positive bias region, the transmission peak appears right at the 
$E_{F}$.

From the above observations, we describe a rectification mechanism for 
the A:T basepair (Fig.3(a)). In 
the negative bias region, the HOMO level for A:T$_{NP}$ and the LUMO level 
for A:T$_{P}$ do not enter into the integrating region (illustrated by the 
shaded region in Fig.3(a)). However, in the positive bias region, low-energy levels,
namely, HOMO level of A:T$_{NP}$ and the LUMO level of A:T$_{P}$ enter the integration 
region and thus we would expect an increase in the current signal for both 
A:T$_{NP}$ and A:T$_{P}$ for this bias polarity. Interestingly however, due to 
the weak transmission (indicated by the dotted lines in Fig. 3a) associated with 
the HOMO level for A:T$_{NP}$ in the positive bias region, the resulting 
current would be quite small. Thus, we note that, only in the A:T$_{P}$ system, 
we will observe an asymmetric rectifying I-V profile with significant current 
in the positive bias polarity and negligible electron flow in the reverse polarity
of applied bias. 

On analyzing the T(E) data for the A:A systems, we find that using the 
reversible equilibrium between A:A$_{NP}$ and A:A$_{P}$, the A:A basepair 
can be used as a molecular switch at zero bias. At equilibrium (zero bias), 
the HOMO level of A:A$_{NP}$ (Fig.2(e)) is 1.05 eV below the 
$E_{F}$, whereas the LUMO level of A:A$_{P}$ (Fig.2(f)) occurs right at the 
$E_{F}$. This is quite significant, as even at zero bias, if the pH of the system 
is changed from neutral to acidic, we would see a switching behavior which 
is reversible. From the T(E) data we also find that 
this switching action can be observed in a very small applied bias window as the 
transmission of the LUMO level of A:A$_{P}$ decreases rapidly under the 
influence of an external bias in both the directions. We depict this scenario 
schematically in Fig. 3(b). The mechanism of the switching behavior can be 
explained by examining the 
HOMO and LUMO orbitals of A:A$_{NP}$ and A:A$_{P}$, respectively. The 
same are plotted in Fig.4. We find that the HOMO of A:A$_{NP}$ is not 
delocalized over the entire basepair, rather is localized 
on the individual Adenine bases, whereas, the LUMO of A:A$_{P}$ is delocalized 
over the entire basepair. Note that, such large delocalization characteristics 
of A:A$_{P}$ 
stems from the $\pi$-$\pi$ interactions between the Adenine bases. 
In a recent experimental study by Seferos {\it et al.} 
the strength of $\pi$-$\pi$ coupling versus conjugated coupling was 
studied and it was found that the $\pi$-$\pi$ coupling improved the 
conduction due to strong {\it through space} electrostatic interactions. This 
is the same reason why A:A$_{P}$ shows stronger conducting profile in
comparison to A:A$_{NP}$\cite{PNAS,lakshmi}. Also, the fact that the 
switching interactions is controlled due to the interplay between weak 
interactions, namely, H-bonding and $\pi$-$\pi$ interactions, promise 
technological applications with easy tunability. Our findings thus suggest that
the noncanonical A:A systems are prospective candidates for 
switching action under very small pH change to the overall system. 

\begin{figure}
\includegraphics[scale=0.09,angle=0]{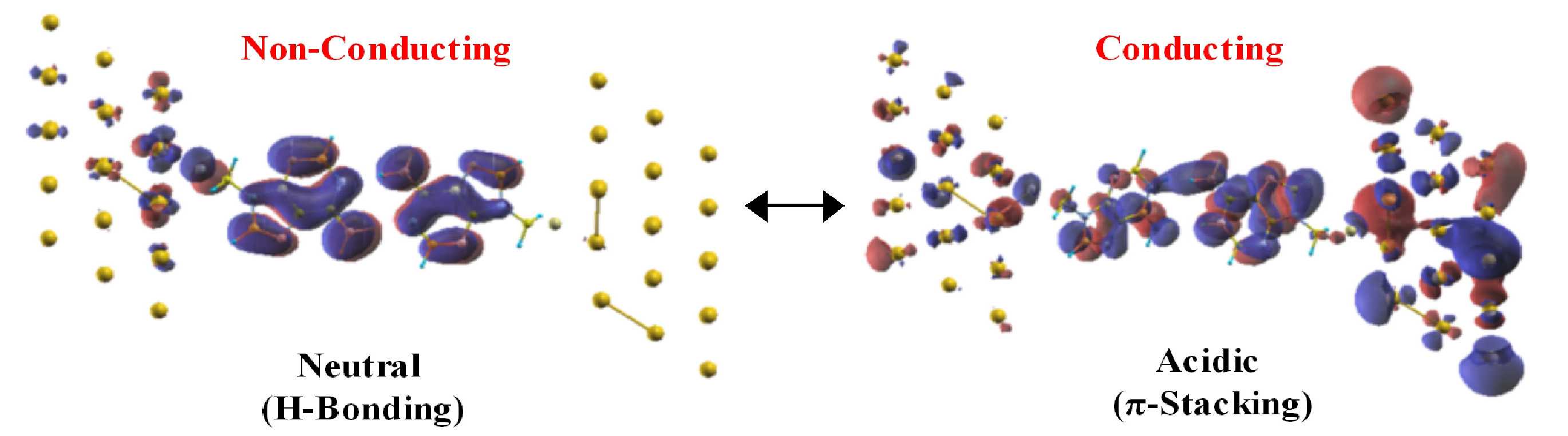}
\caption{(Color online) Orbital plots of the basepairs; HOMO of A:A$_{NP}$ (left)
and LUMO of A:A$_{P}$ (right).} 
\end{figure}

In conclusion, we find that transport characteristics of the A:T and the 
G:C basepairs vary significantly from each other. The H-bonds in A:T are 
energetically very different from those in G:C; while A:T favors efficient 
transport, G:C does not. We find that under mild acidic conditions, the A:T basepairs 
can be used as efficient molecular rectifiers; the action of which can be enhanced by 
tuning the pH of the system. Also we find that the non-canonical A:A basepair can 
be used as an efficient molecular switch at zero bias, where the switching behavior 
is controlled by an interplay between weak intermolecular H-bonding and 
$\pi$-$\pi$ interactions.

S. S. M. thanks CSIR for financial support. S. K. P. acknowledges the DST, 
Government of India for financial support.

\end{document}